\newcommand{\D}[1]{\text{d}#1}
\newcommand{\sign}{{\rm sgn}\,}
\newcommand{\tr}{{\rm tr}\,}
\documentclass[%
 reprint,
superscriptaddress,
 amsmath,amssymb,
 aps,
]{revtex4-2}

\def\XXint#1#2#3{{\setbox0=\hbox{$#1{#2#3}{\int}$}
     \vcenter{\hbox{$#2#3$}}\kern-.5\wd0}}

\usepackage{amsthm} 
\usepackage{tikz}
\usepackage{pgfplots}
\usepackage{color}

\def\doi{http://dx.doi.org/}
\usepackage{braket}
\definecolor{OliveGreen}{RGB}{85,107,47}
\definecolor{NavyBlue}{RGB}{0,0,128}
\usepackage[colorlinks,bookmarks=false,citecolor=NavyBlue,linkcolor=OliveGreen,urlcolor=blue]{hyperref}
\usepackage{feynmf,tikz, relsize}
\newcommand{\bw}{\begin{widetext}}
\newcommand{\ew}{\end{widetext}}
\newcommand{\be}{\begin{equation}}
\newcommand{\ee}{\end{equation}}
\newcommand{\bea}{\begin{eqnarray}}
\newcommand{\eea}{\end{eqnarray}}
\def\nn{\nonumber\\}

\usepackage{graphicx}
\usepackage{dcolumn}
\usepackage{bm}

\pgfplotsset{compat=1.17}
\begin{document}


\title{Duality between weak and strong interactions in quantum gases}

\author{Etienne Granet}
\affiliation{Rudolf Peierls Centre for Theoretical Physics, Clarendon Laboratory, Oxford OX1 3PU, UK}
\author{Bruno Bertini}
\affiliation{Rudolf Peierls Centre for Theoretical Physics, Clarendon Laboratory, Oxford OX1 3PU, UK}
\author{Fabian H.L. Essler}
\affiliation{Rudolf Peierls Centre for Theoretical Physics, Clarendon Laboratory, Oxford OX1 3PU, UK}

\date{\today}

\begin{abstract}
In one dimensional quantum gases there is a well known ``duality'' between hard core bosons and non-interacting fermions. However, at the field theory level, no exact duality connecting strongly interacting bosons to weakly interacting fermions is known. Here we propose a solution to this long standing problem. Our derivation relies on regularizing the only point-like interaction between fermions in 1D that induces a discontinuity in the wave function proportional to its derivative. In contrast to all known regularisations our potential is weak for small interaction strengths. Crucially, this allows one to apply standard methods of diagrammatic perturbation theory to strongly interacting bosons. As a first application we compute the
finite temperature spectral function of the Cheon-Shigehara model, the fermionic model dual to the celebrated Lieb--Liniger model.\\

\end{abstract}

\maketitle


Duality is an important concept in physics that refers to alternative descriptions of the same physical situation. This is particularly useful in cases where duality relates strongly interacting theories to weakly interacting ones. A prototypical example is the Bose-Fermi mapping in 1+1 dimensional field theories with point-like interactions \cite{giradeau1960relationship,cheon1998realizing,cheon1999fermion} which relates the Lieb--Liniger (LL) \cite{lieb1963exact} and Cheon--Shigehara (CS) \cite{cheon1999fermion} models. Given that the LL model of strongly interacting bosons has been at the heart of numerous experimental discoveries over the last two decades, see e.g. \cite{kinoshita2004observation,paredes2004tonks,kinoshita2006quantum,haller2010pinning,tang2018thermalization,meinert2017bloch,malvania2020generalized,Wilson2020observation, schemmer2019generalized, bouchoule2021generalized}, one would expect the Bose-Fermi duality to provide a very useful tool for understanding the observed behaviours. However, the CS form of the fermionic interaction potential does not allow for a perturbative analysis and masks the fact that there exists a weakly coupled regime corresponding to strong interactions between the LL bosons. There have been previous attempts to reformulate the CS interaction in order to clearly exhibit this weakly coupled regime, but these either violate the non-perturbative duality \cite{sen1999perturbation,sen2003the} (even though they allow for first order perturbative calculations \cite{Sykes2008spatial,deuar2009nonlocal} as well as Hartree-Fock and random-phase approximations and low-energy effectivel field theories \cite{brand2005dynamics,Brand2006polarizability,Cherny2009dynamic,valiente2015effective,valiente2020bose,valiente2021universal,pastukhov2020ground}) or cannot be formulated in second quantization \cite{girardeau2004theory}. In the following we present a non-perturbative reformulation of the CS model that makes the existence of a weak coupling regime manifest and allows the full machinery of many-particle perturbation theory to be applied.

At the heart of our approach is the much studied 
problem of point-like interactions in quantum 
mechanics. It is well-known that bosons interacting in 1D via a point-like potential have a wave-function that is continuous when two bosons coincide, but with a discontinuous derivative \cite{albeverio1988solvable}. For two particles the relative motion is described by the textbook Schr\"odinger equation
\begin{equation}\label{eqdelta}
\phi''(x)+k^2\phi(x)=2\gamma \delta(x)\phi(x)\,,
\end{equation}
with $\gamma$ a coupling parameter. Integrating over a small interval $-\epsilon<x<\epsilon$ around zero one obtains the condition $\phi'(0^+)-\phi'(0^-)=2\gamma \phi(0)$. However, this manipulation should be considered as heuristic, since the product of a distribution with a non-smooth function is ill-defined. The proper way of analyzing \eqref{eqdelta} and deriving the discontinuity of $\phi'(x)$ at zero is by \emph{regularizing} the $\delta$ distribution, namely by replacing it with a smooth function $\delta_a(x)$ satisfying $\int \delta_a(x)\D{x}=1$ and $\lim_{a\to 0}\, \delta_a(x)=0\text{ for }x\neq 0$. Then solution to (\ref{eqdelta}) is  obtained as the limit $a\to 0$ of the even solution $\phi_a(x)$ to the corresponding regular Schr\"odinger equation. This interpretation is physically meaningful as a point-like potential is merely an approximation of a regular potential whose range is much smaller than the wave length of the bosons. 

The fermionic counterpart of this problem is less known and more subtle. For fermions, the only parity symmetric point-like potential induces a discontinuity \textit{in the wave function itself} (see below) \cite{seba1986generalized,albeverio1998symmetries,albeverio1988solvable} \footnote{Apart from totally reflective potentials that make  the regions $x>0$ and $x<0$ independent.}. But in contrast to the bosonic case, the interpretation of this point-like potential in terms of (derivatives of) $\delta$-functions is problematic \cite{seba1986some}. 
The heuristic analog of the Schr\"odinger equation \eqref{eqdelta} would be \cite{kurasov1996distribution,kurasov1998finite,albeverio1998symmetries}
\begin{equation}\label{eqdelta2}
\psi''(x)+k^2\psi(x)=2\beta \partial_x[\delta(x)\partial_x]\psi(x)\,,
\end{equation}
where $\beta$ parametrizes the interaction strength. Indeed, by integrating $x$ over the interval $-\epsilon<x<y$ and then $y$ over $-\epsilon<y<\epsilon$ and applying the usual rules for the $\delta$ distribution yields $\psi(0^+)-\psi(0^-)=2\beta \psi'(0)$ \cite{grosse2004exact}. Through the Girardeau mapping  $\phi(x)=\sign(x)\psi(x)$ \cite{giradeau1960relationship,cheon1999fermion} the fermionic point-like interaction \eqref{eqdelta2} is dual to the bosonic one \eqref{eqdelta} with $\beta=\frac{1}{\gamma}$.  However, this manipulation of $\delta$'s to obtain the jump condition is again problematic and requires a more careful analysis. In contrast to \eqref{eqdelta}, this time a simple regularisation of the delta function does not suffice. Indeed, replacing $\delta(x)$ by a smooth potential $\delta_a(x)$ in \eqref{eqdelta2} \textit{does not} yield the expected discontinuity in the limit $a\to 0$ and one obtains instead a continuous wave function~\footnote{See the Supplemental Material for: (i) a brief proof of the absence of discontinuity in Eq.~\eqref{eqdelta2} with the replacement $\delta(x)\to \delta_a(x)$; the precise definition of the Feynman rules for the Hamiltonian~\eqref{eq:FermionicH}; (iii) an explicit proof of~\eqref{eq:SelfEnR}.}. One way of making the above manipulations well-defined is to consider a generalization of distributions to discontinuous test functions~\cite{kurasov1998finite,seba1986some,albeverio1998symmetries,kurasov1996distribution,kurasov1998finite}. Other generalizations of distributions were also considered \cite{valiente2020bose}. Although mathematically sound, these generalizations suffer from a lack of physical meaning: one loses the interpretation of the point-like interaction as an approximation of a very short range smooth potential. Moreover, the meaning of \eqref{eqdelta2} in second quantization becomes unclear.


The physical and operational definition of this point-like interaction in terms of a regularization is thus a non-trivial problem. It requires the construction of a smooth potential 
$V_a(x)$ such that the odd solution $\psi_a(x)$ to the Schr\"odinger equation
\begin{equation}\label{shord4}
\psi_a''(x)+k^2\psi_a(x)=V_a(x)\psi_a(x)\,,
\end{equation}
reduces in the limit $a\to 0$ to 
\begin{align}\label{shrod5}
 \psi''(x)+k^2\psi(x)&=0\ \text{for }x\neq 0\ ,\hfill\nn
\psi'(0^+)-\psi'(0^-)&=0\ ,\nn
\psi(0^+)-\psi(0^-)&=2\beta\psi'(0)\ .
\end{align}
Solutions to this problem involving non-Hermitian potentials, non-local potentials or pseudo-potentials were proposed \cite{carreau1993four,chernoff1993a,roman1996the,albeverio2000approximation,golovaty2018two}. The first solution in terms of a Hermitian, regular potential was obtained by Cheon and Shigehara in Ref.~\cite{cheon1998realizing} and is of the form
\be
V_a^{\rm CS}(x)=\left[\frac{1}{\beta}-\frac{1}{a}\right]\big(\delta(x+a)+\delta(x-a)\big).
\ee
Here the $\delta$ functions can be regularized on a scale that is small compared to $a$ \cite{exner2001potential,zolotaryuk2017families}. Crucially this potential describes \textit{strong interactions} between two fermions for any value of $\beta$. This is in spite of the fact that for small $\beta$ it results in wave functions that are close to those of free fermions. While this formulation allowed CS to establish a duality between a system of interacting fermions and the LL model, it obscured the fact that strongly interacting bosons are dual to \emph{weakly} interacting fermions. Moreover, by its very nature it precluded any kind of perturbative calculation. 
The key aspect of our work is the construction of a smooth potential $V_a(x)$ that gives rise to \eqref{shrod5} while being weak for small $\beta$.

\emph{A smooth weakly coupled potential for fermions.} 
For a coupling strength $\beta> 0$ and a regularization parameter $a>0$, we define the following smooth potential
\begin{equation}\label{potential}
V_{a,\beta}(x)= \frac{\beta\sigma''_a(x)}{x+\beta \sigma_a(x)}\,,
\end{equation}
with $\sigma_a(x)\equiv \sigma(x/a)$ where $\sigma(x)$ is any odd regular function that satisfies
\begin{equation}\label{sigma}
\begin{aligned}
&\underset{x\to\infty}{\lim}\, \sigma(x)=1\,,& &\underset{x\to\infty}{\lim}\, x^2\sigma''(x)=0\\
&\sigma'(0)> 0\,,& &\forall x,\,\sigma'(x)\geq 0\,.
\end{aligned}
\end{equation}
For example, one can choose $\sigma_a(x)=\tanh {x}/{a}$. Let $\psi_{a,\beta}(x)$ be the odd solution to the Schr\"odinger equation
\begin{equation}\label{eqv}
\psi_{a,\beta}''(x)+k^2\psi_{a,\beta}(x)=V_{a,\beta}(x)\psi_{a,\beta}(x)\,,
\end{equation}
with a fixed boundary condition $\psi_{a,\beta}(1)=1$. The key result of this Letter is that in the limit $a\to 0$ at fixed $\beta>0$, the wave function $\psi_{a,\beta}(x)$ of the potential \eqref{potential} satisfies \eqref{shrod5}. 

We now briefly sketch the proof of this statement. The idea is to treat $k^2\psi_{a,\beta}(x)$ in \eqref{eqv} as an inhomogeneous term in the homogeneous equation obtained for $k=0$. The even $\phi_{a,\beta}^+$ and odd $\phi_{a,\beta}^-$ independent solutions of this homogeneous equation are
\begin{equation}
\begin{aligned}
    &\phi_{a,\beta}^-(x)=x+\beta\sigma_a(x)\\
    &\phi_{a,\beta}^+(x)=\frac{1}{1+\beta\sigma_a'(x)}\\
    &+(x+\beta \sigma_a(x))\int_{0}^x \D{y}\frac{\beta\sigma_a''(y)}{(y+\beta\sigma_a(y))(1+\beta\sigma_a'(y))^2}\,.
\end{aligned}
\end{equation}
Applying the method of variation of parameters, one obtains the following self-consistency condition for the odd solution to \eqref{eqv} for $k\neq 0$
\begin{align}\label{self}
\psi_{a,\beta}(x)&=k^2\sum_{\sigma=\pm}\sigma \phi_{a,\beta}^\sigma(x)\int_{0}^x \D{y} \psi_{a,\beta}(y)\phi_{a,\beta}^{-\sigma}(y)\nn
&+A \phi_{a,\beta}^-(x)\ ,
\end{align}
where $A$ is an integration constant. An analysis of $\phi_{a,\beta}^+(x)$ based on the assumptions \eqref{sigma} shows that it can be bounded independently of $a$ and $x\in[-1,1]$ and that for $x\neq 0$
\begin{equation}
    \underset{a\to 0}{\lim}\, \phi_{a,\beta}^+(x)=-\frac{|x|}{\beta}\,.
\end{equation}
From this and \eqref{self}, Gr\"onwall's inequality~\cite{ames1997inequalities} implies then that $\psi_{a,\beta}(x)$ itself can be bounded independently of $a$ and $x\in[-1,1]$. This allows us to commute the limit $a\to 0$ and the integrations in \eqref{self}. The resulting self-consistency equation for ${\lim}_{a\to 0}\,\psi_{a,\beta}(x)$ establishes then that it satisfies \eqref{shrod5}. 



\emph{N-particle Cheon--Shigehara gas.} 
Following Refs~\cite{gaudin1983la,cheon1999fermion} the above $2$-particle result can be readily extended to a gas of $N$ fermions 
with Hamiltonian
\begin{equation}\label{hams}
H^{\rm f}_{a,\beta}=-\sum_{j=1}^N \partial_{x_j}^2+ 2\sum_{j<k}V_{a,\beta}(x_j-x_k)\,,
\end{equation}
which from now on we will refer to as CS gas. The reasoning goes as follows. First one observes that, since the eigenstates $\psi^{\rm f}_{a,\beta}(x_1,...,x_N)$ are antisymmetric functions of $x_1,...,x_N$ it suffices to know them on
$D=\{x_1<...<x_N\}$. Second one notes that, because the potential \eqref{potential} satisfies \eqref{shrod5} when $a\to 0$, the many-body wave-function fulfils $\sum_j \partial_{x_j}^2\psi^{\rm f}_{0,\beta}=0$ inside $D$, and obeys the following conditions at the boundary of $D$
\begin{equation}
\begin{aligned}
\psi^{\rm f}_{0,\beta}\Big|_{x_j=x_{j+1}^-}
&=-\frac{\beta}{2}[\partial_{x_{j+1}}-\partial_{x_j}]\psi^{\rm f}_{0,\beta}\Big|_{x_j=x_{j+1}^-}\,.\\
\end{aligned}
\end{equation}
Finally, performing the Girardeau mapping \cite{giradeau1960relationship,yukalov2005fermi}\footnote{In second quantization, this mapping is the Jordan-Wigner transformation.}
\begin{equation}
\!\!\psi^{\rm b}_{2/\beta}(x_1,...,x_N)\!=\psi^{\rm f}_{0,\beta}(x_1,...,x_N)\prod_{j<k}\sign(x_j-x_k),
\end{equation}
one finds that the function $\psi^{\rm b}_{2/\beta}$ exactly satisfies the conditions of the LL eigenstates at the boundary of $D$ \cite{gaudin1983la}. This establishes that when $a\to 0$ at fixed $\beta$, $H^{\rm f}_{a,\beta}$ is equivalent to the LL Hamiltonian 
\begin{equation}\label{ll}
H^{\rm b}_c=-\sum_{j=1}^N \partial_{x_j}^2+ 2c\sum_{j<k}\delta(x_j-x_k)\,,
\end{equation}
with $\beta=2/c$.
Having established that (\ref{hams}) provides a dual description to (\ref{ll}) we now show that our formulation allows one to carry our perturbative calculation in the large-$c$ limit that are in agreement with exact results. When ${\beta=0}$, the energy levels of the free fermion Hamiltonian \eqref{hams} on a ring of size $L$ are given by $\sum_{\lambda_i\in\pmb{\lambda}}\lambda_i^2$ with $\pmb{\lambda}$ any subset of $\{{2\pi n}/{L},n\in\mathbb{Z}\}$ with $N$ elements. In the thermodynamic limit $L\to\infty$, they are parametrized by a particle density ${0\leq\rho(\lambda)\leq {1}/{(2\pi)}}$. Let us now fix such a state at $\beta=0$ through its particle density $\rho$ and compute perturbatively in $\beta$ the energy levels of \eqref{hams} at fixed $a$. The energy per site $e_a(\beta)$ can be written as
\begin{equation}
e_a(\beta)=e^{(0)}_a(\beta)+e^{(1)}_a(\beta)+e^{(2)}_a(\beta)+\mathcal{O}(\beta^3)\,,
\end{equation}
where the successive orders in perturbation theory are
\begin{align}
\label{shrper}
e^{(0)}_a(\beta)&=\int_{-\infty}^\infty \D{\lambda} \rho(\lambda) \lambda^2\ ,\nn
e^{(1)}_a(\beta)&=\int_{-\infty}^\infty \D{\lambda}\D{\mu}\rho(\lambda)\rho(\mu)[\hat{V}_{a,\beta}(0)-\hat{V}_{a,\beta}(\lambda-\mu)]\ ,\nn
e^{(2)}_a(\beta)&=\pi\int_{-\infty}^\infty \D{\lambda}\D{\mu}\D{\nu}\rho(\lambda)\rho(\mu) \rho_h(\lambda+\nu)\rho_h(\mu-\nu)\nn
&\qquad\times\ \frac{[\hat{V}_{a,\beta}(\lambda-\mu+\nu)-\hat{V}_{a,\beta}(\nu)]^2}{\nu(\mu-\lambda-\nu)}\ .
\end{align}
Here $\hat{V}_{a,\beta}(\lambda)=\int_{-\infty}^\infty \D{x} V_{a,\beta}(x)e^{i\lambda x}$ and $\rho_h(\lambda)={1}/({2\pi})-\rho(\lambda)$ is the hole density. Expanding the potential \eqref{potential} in $\beta$ at fixed $a$ and considering $a\to 0$ afterwards, we have up to $\mathcal{O}(a)+\mathcal{O}(\beta^3)$ corrections 
\begin{align}
e^{(1)}_a(\beta)=&2\beta \mathcal{D}\mathcal{E}\left[-1+\frac{\beta}{2a}\int_{-\infty}^\infty \D{x} \sigma'(x)^2 \right]\,,\\
e^{(2)}_a(\beta)=&3\beta \mathcal{D}\mathcal{E}\left[1-\frac{\beta}{3a}\int_{-\infty}^\infty \frac{\D{\omega} }{2\pi} [\widehat{\sigma'}(\omega)]^2\right]\ .
\end{align}
Here $\mathcal{D}=\int \D{\lambda} \rho(\lambda)$ and $\mathcal{E}=\int \D{\lambda} \rho(\lambda)\lambda^2$ are respectively the particle density and the (unperturbed) energy density of the macro state parametrized by $\rho(\lambda)$. 
We observe that the first and second order contributions are both divergent in $1/a$, but remarkably, using Parseval's identity we find that their sum is 
in fact finite
\begin{equation}\label{rese}
e_a(\beta)=\left(1-2\beta \mathcal{D}+3\beta^2\mathcal{D}^2 \right)\mathcal{E}+\mathcal{O}(a)+\mathcal{O}(\beta^3)\ .
\end{equation}
The compensation is due to the specific form of the potential \eqref{potential}. 
The expression (\ref{rese}) agrees with the exact Bethe ansatz result for the LL model with $\beta={2}/{c}$ at order $1/c^2$. Our calculation shows that all energy levels of \eqref{hams} can be computed perturbatively in $\beta$ at fixed $a$ and then the regulator $a$ can be sent to $0$ order by order in $\beta$ to order $\beta^2$.

Note that the result \eqref{rese} allows one to study the thermodynamics of the gas \eqref{hams} up to order $1/c^2$. Indeed, one can calculate the free energy at temperature $T$, 
\be
F_{a,\beta}=\tr\left[e^{-H^{\rm f}_{a,\beta}/T}\right],
\ee
expanding the trace in terms of the non-interacting basis. One can then use \eqref{rese} and proceed as in the Thermodynamic Bethe Ansatz (TBA) treatment~\cite{korepin1993quantum, yang1969thermodynamics}. For example, in this way one finds that, up to order $1/c^2$, the thermal energy density is given by the expression \eqref{rese} with $\rho$ being the thermal root density satisfying the Yang-Yang equation expanded at order $c^{-2}$.



\emph{Cheon--Shigehara Field theory.} 
The Hamiltonian \eqref{hams} (on a ring of length $L$) can be expressed in second quantization as 
\be
\!\!{\mathcal H}^{\rm f}_{a,\beta} \!=\!  \sum_{p} (p^2-\mu) \psi_p^\dag \psi_{p}+\!\sum_{{\boldsymbol p}} W_{a,\beta}({\boldsymbol p}) \psi_{p_1}^\dag \psi_{p_2}^\dag\psi_{p_3}\psi_{p_4}\ ,
\label{eq:FermionicH}
\ee
where $\psi^\dag_p$ and $\psi_p$ are canonical Fermi fields in momentum space, 
and we have introduced the short hand notation ${\boldsymbol p} \equiv (p_1,\ldots,p_4)$.
The interaction vertex in~\eqref{eq:FermionicH} is given by
\begin{align}
W_{a,\beta}({\boldsymbol p})  = & \frac{1}{4L}\delta_{p_1+p_2-p_3-p_4,0}\notag\\
 &\times \sum_{P,Q\in S_2} \!\!\!\!{\rm sgn}(PQ) \hat V_{a,\beta}(p_{P(1)}-p_{Q(1)+2})\,,
\end{align}
where $S_2$ is the group of permutations of two elements. 

\begin{figure}[b]
 \includegraphics[scale=0.115]{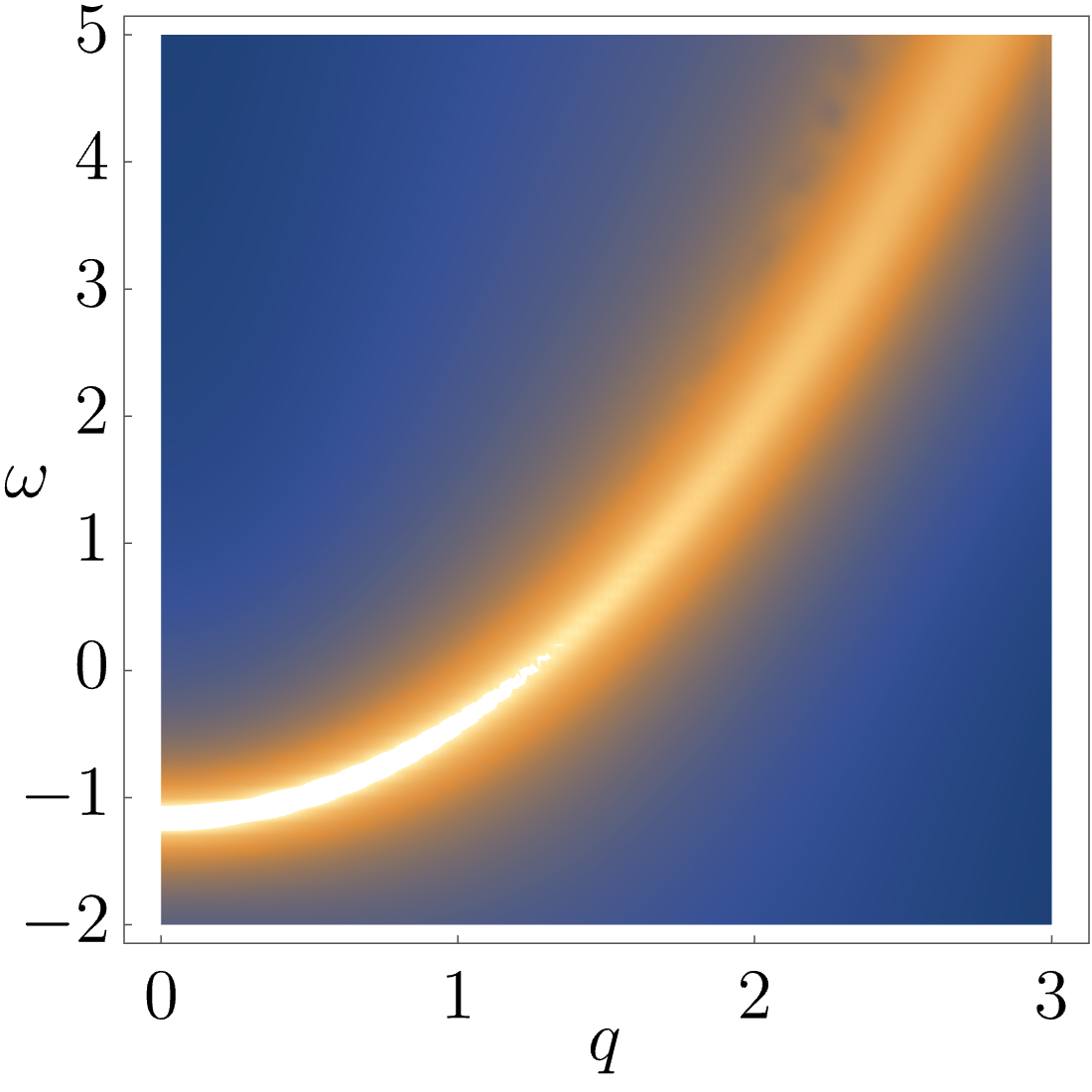}
 \includegraphics[scale=0.115]{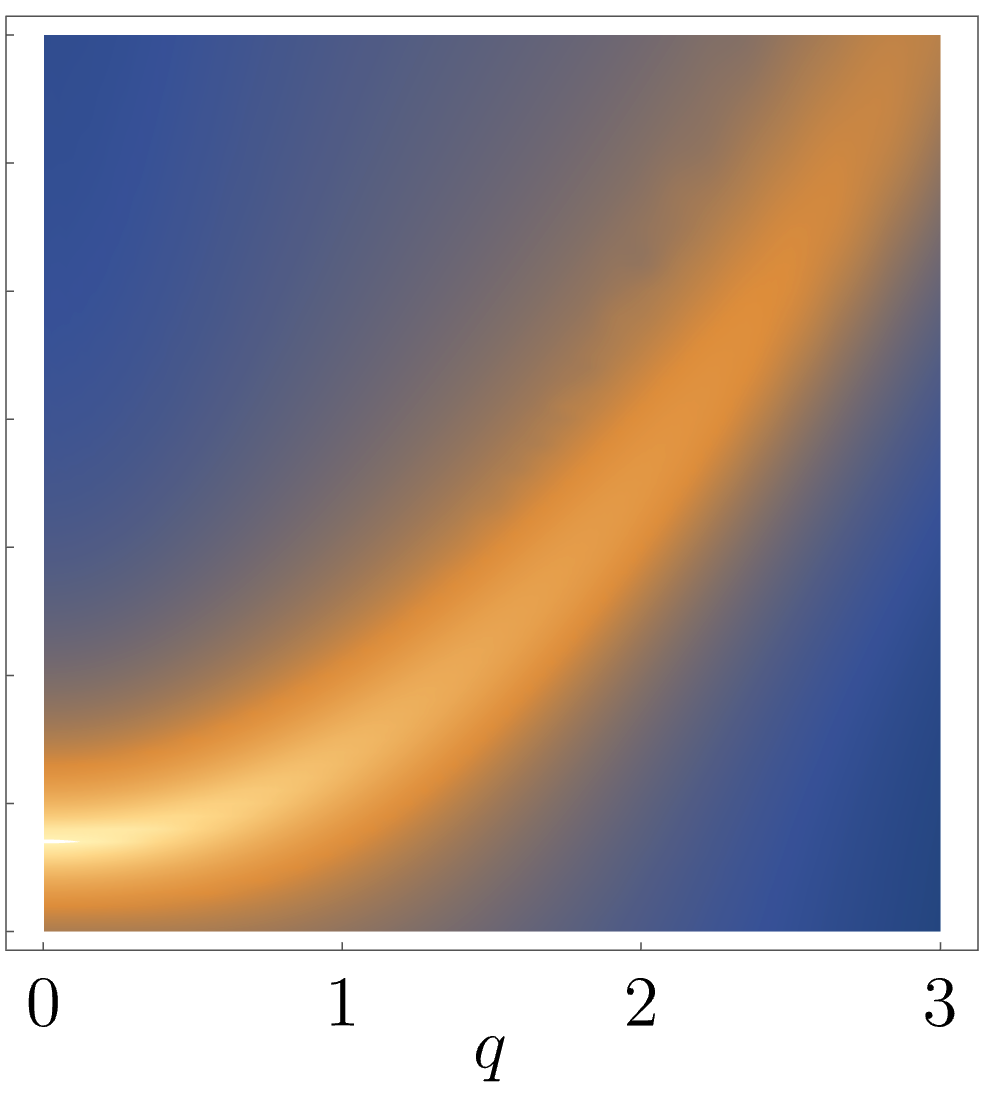}
\caption{Spectral function of the CS gas $A(\omega,q)$ for $\beta=2/c=0.5$ (left) and $\beta=1$ (right) in an equilibrium state at temperature $T=1$ and chemical potential $\mu=1$. The color scale is the same for both plots. The free fermion $\beta=0$ spectral function is $2\pi \delta(\omega-q^2+\mu)$.}
\label{fig:SF}
\end{figure}

For small $\beta$ the theory \eqref{eq:FermionicH} can be analyzed using standard diagrammatic perturbation theory. Let us consider in particular the thermal propagator  
\be
G(\tau,k) = -\frac{\textrm{tr}\bigl[T_\tau\bigl[ e^{\tau {\mathcal H}^{\rm f}_{a,\beta}} \psi_k e^{-\tau {\mathcal H}^{\rm f}_{a,\beta}} \psi^\dag_k\bigr] e^{-{\mathcal H}^{\rm f}_{a,\beta}/T}\bigr]}{\textrm{tr}\bigl[e^{-{\mathcal H}^{\rm f}_{a,\beta}/T}\bigr]}.
\label{eq:thermalpropagator}
\ee
The usual procedure (see, e.g.,~\cite{bruus2004many}) is to exploit the anti-periodicity of \eqref{eq:thermalpropagator} for $\tau \mapsto \tau+1/T$, and consider its Fourier coefficients, denoted by ${G}(\omega_n,k)$, where $\omega_n = {2\pi} T(n+1/2)$ are called Matsubara frequencies. These coefficients can be written in the following Dyson form  
\be
{G}(\omega_n,k) = \frac{1}{i\omega_n-k^2+\mu-\Sigma(\omega_n,k)} \ ,
\label{eq:Gomega}
\ee
where the proper self energy, $\Sigma(\omega_n,k)$, is defined as the sum of all irreducible Feynman diagrams with two amputated legs~\cite{bruus2004many}. The self energy encodes all information about the thermodynamics of the system 
as well as very relevant information about its dynamics. Indeed, it can be used to determine both the free energy and the spectral function~\cite{kadanoff1962quantum}. Specifically, the latter is expressed as $A(\omega,k)=-2 {\rm Im}(G^{R}(\omega,k))$, where the Fourier transform of the retarded Green's function $G^{R}(\omega,k)$ is obtained by performing the analytic continuation $i \omega_n\mapsto \omega + i 0^+$ in \eqref{eq:Gomega}.  
For the theory \eqref{eq:FermionicH}, considering contributions up to order $\beta^2$, we find    
\begin{widetext}
\be
\Sigma(\omega_n, k)=\raisebox{-17pt}{\scalebox{.8}{
\begin{fmffile}{firstorder}
\begin{fmfgraph*}(120,45)
\fmfleft{i1}
\fmftop{k1}
\fmfright{o1}
\fmf{scalar,label=${\omega_{n}},,{k}$}{i1,v1}
\fmf{scalar,label=${\omega_{n}},,{k}$}{v1,o1}
\fmf{fermion,label=${\omega_{m}},,{q}$}{v1,v1}
\end{fmfgraph*}
\end{fmffile}}}+\raisebox{-17pt}{\scalebox{0.8}{
\begin{fmffile}{secondorder2}
\begin{fmfgraph*}(140,45)
\fmfleft{i1}
\fmfright{o1}
\fmf{scalar,label=${\omega_{n}},,{k}$}{i1,v1}
\fmf{fermion,label=${\omega_{m_2}},,{q_2}$,l.side=left, l.dist=250, tension=0.3}{v1,v2}
\fmf{fermion,label=${\omega_{m_1}},,{q_1}$,l.side=right, tension=0.3}{v2,v1}
\fmf{fermion,label=${\omega_{m_3}},,{q_3}$,l.side=right, l.dist=100, tension=0.3}{v1,v2}
\fmf{scalar,label=${\omega_{n}},,{k}$}{v2,o1}
\end{fmfgraph*}
\end{fmffile}}}+\!\!\!\raisebox{2pt}{\scalebox{0.8}{
\begin{fmffile}{secondorder1}
\begin{fmfgraph*}(130,85)
\fmfleft{i1,i2,i3}
\fmfright{o1,o2,o3}
\fmf{scalar,label=${\omega_{n}},,{k}$}{i1,v1}
\fmf{scalar,label=${\omega_{n}},,{k}$}{v1,o1}
\fmf{phantom}{i2,v2,o2}
\fmf{phantom}{i3,v3,o3}
\fmffreeze
\fmf{fermion,label=${\omega_{m_1}},,{q_1}$,tension=1.8,right}{v1,v2}
\fmf{fermion,tension=1.8,right}{v2,v3,v2}
\fmf{fermion,label=${\omega_{m_3}},,{q_3}$,tension=1.8,right}{v2,v1}
\fmfv{label=${\omega_{m_2}},,{q_2}$}{v3}
\end{fmfgraph*}
\end{fmffile}}}
+\mathcal{O}(\beta^3)
\ee
where the incoming and outgoing legs (dashed lines) are amputated~\cite{Note1}. 
Remarkably, evaluating these diagrams we find that, in analogy to what happens for \eqref{rese}, the $1/a$ divergences in the second order contributions compensate and the final result does not require further regularization~\cite{Note1}. Specifically, in the thermodynamic limit we have  
\begin{align}
\Sigma(\omega_n, k) =& -2 \beta (A_2 + A_0 k^2)-\frac{2 \beta^2}{T} \int\frac{{\rm d} q}{2\pi} (A_2 + A_0 q^2) (k-q)^2 n(q) (1-n(q))\notag\\
& +2 \beta^2 \int \frac{{\rm d} q_2}{2\pi}\frac{{\rm d} q_3}{2\pi} \frac{{((k-q_3)^2-(q_2-q_3)^2)^2}}{i \omega_n +q_2^2-q_3^2-\bar q_4^2+\mu}  (n(q_3)n(\bar q_4)-n(q_2)n(\bar q_4)-n(q_2)n(q_3))\notag\\
& - i \beta^2 \int \frac{{\rm d} q_2}{2\pi} \sqrt{2(i \omega_n-k^2+\mu)+(k-q_2)^2} (k-q_2)^2 n(q_2)\ ,
\label{eq:SelfEnR}
\end{align}
\end{widetext}
where we chose the branch cut of the square root to lie along the positive real axis, $\bar q_4=k+q_2-q_3$, and 
\begin{align}
n(p) & = \frac{1}{1+e^{(p^2-\mu)/T}}\ ,\quad
A_m=\int \frac{{\rm d} p}{2 \pi}\, p^m\ n(p)\ .
\end{align}
To the best of our knowledge \eqref{eq:SelfEnR} is the first calculation of the self energy in the CS model at second order in $\beta$ --- where interactions generate a non-vanishing imaginary part --- and represents our second main result. We verified that (i) the spectral function $A(\omega,k)$ obtained from \eqref{eq:SelfEnR} fulfils the exact sum rule $\int {\rm d}\omega/(2\pi)\, A(\omega,k)=1$; (ii) the internal energy per volume computed using \eqref{eq:SelfEnR} agrees up to order $1/c^2$ with the exact result for the LL model~\cite{korepin1993quantum, yang1969thermodynamics}. 
The non-trivial effects of the interactions are best appreciated by considering the spectral function, \emph{cf.} Fig.~\ref{fig:SF}. We see that as a result of the interactions the fermions created by $ \psi^\dag_k$ acquire a finite life-time and the dispersion gets renormalized. 
We note that the appearance of a finite lifetime is not in contradiction with the integrability of the theory \eqref{eq:FermionicH} because the integrability-protected stable quasiparticles differ from the fermions created by $\psi^\dag_k$ for finite $c$ \cite{creamer1980quantum}.

\emph{Discussion.} In this Letter we presented a one dimensional quantum mechanical potential that induces a discontinuity in the wave function proportional to its derivative and, at the same time, can be expanded perturbatively for small but finite interaction strengths. This addresses the long-standing problem of how to best regularize point-like interactions in quantum mechanics. We used this potential to obtain a reformulation of the Cheon-Shigehara gas, the fermionic theory dual to the Lieb-Liniger model, that makes it manifest that strongly interacting bosons correspond to weakly coupled fermions. Our results open the door to the systematic analysis of strongly interacting bosons away from the hard-core limit by means of perturbation theory, going considerably beyond the current state of the art. As a first application we have obtained the spectral function of the CS model at order $1/c^2$, displaying the previously inaccessible broadening shown in Fig.~\ref{fig:SF}.

Our work can be generalized in a number of ways. It can for example be directly extended to treat Bose-Fermi dualities in multi-species systems \cite{ohya2021}, allowing one to access cases of high experimental relevance~\cite{kinoshita2004observation,paredes2004tonks,kinoshita2006quantum,haller2010pinning,tang2018thermalization,meinert2017bloch,malvania2020generalized,Wilson2020observation, schemmer2019generalized, schweigler2017experimental} that up to now have been treated only in the limit of infinite repulsion~\cite{settino2021exact,minguzzi2002high, vignolo2013universal, rigol2005finite}, via low-energy approximations~\cite{gritsev2007linear, deng2008phase, citro2020spectral, vannieuwkerk2019self, vannieuwkerk2020on}, or in the hydrodynamic regime~\cite{schemmer2019generalized, doyon2017large, bouchoule2021generalized}. Crucially, our method does not rely on integrability and allows one to study \emph{any} such strongly coupled theory.
The case of \textit{attractive} interactions corresponding to $\beta<0$ is not covered by our potential \eqref{potential} and deserves attention. On physical grounds, we  expect a drastic change of potential from small positive to small negative $\beta$ as it corresponds to going from a Tonks-Girardeau to a Super-Tonks-Girardeau gas. 
Applying our results to a strong coupling expansion in the LL model is particularly appealing due to the recent accounts of uniform convergence --- in space and time ---  of the perturbative series for correlation functions both in~\cite{granet2020a} and out of equilibrium~\cite{granet2021systematic}. For instance, this opens the door to a systematic investigation of quantum quenches, explicitly accessing the late time regime where homogeneous systems are expected to locally relax~\cite{essler2016quench} and inhomogeneous ones to follow the predictions of generalised hydrodynamics~\cite{castro2016emergent, bertini2016transport, alba2021generalizedhydrodynamic}. This could potentially lead to \emph{ab initio} derivations of these expectations in the presence of interactions and a full characterisation of the relaxation mechanisms, a task that has currently been accomplished only for a special quantum cellular automaton~\cite{klobas2021exact, klobas2021exactII}. Finally, our work paves the way for establishing the Bose-Fermi mapping at an operatorial level directly in the (regularized) respective field theories. Such a mapping is highly desirable in order to be able to calculate quantities like the boson propagator in the fermionic setting.

{\em Acknowledgments:}
This work has been supported by the EPSRC under grant EP/S020527/1 (EG and FHLE) and the Royal Society through the University Research Fellowship No.\ 201101 (BB).

\bibliography{./bibliography}

\onecolumngrid
\newpage 

\begin{center}
{\large{\bf Supplemental Material for\\
``Realizing discontinuous wave functions perturbatively"}}
\end{center}

Here we report some useful information complementing the main text. In particular
\begin{itemize}
\item[-] In Section~\ref{sec:AD} we present a brief proof of the absence of discontinuity in Eq. (2) of the main text whenever one replaces $\delta(x)$ with a smooth function; 
\item[-] In Section~\ref{sec:FR} we define the Feynman rules for the diagrammatic perturbation theory of the Hamiltonian (22); 
\item[-] In Section~\ref{sec:SR} we present an explicit expression for the proper self energy (27) in terms of the regularised potential (24) and show that it gives (28) in the $a\to 0$ limit;  
\end{itemize}

\section{Absence of discontinuity in (2)}
\label{sec:AD}
In this section we consider the equation
\begin{equation}\label{eqdelta5}
\psi_a''(x)+k^2\psi_a(x)=\beta \sigma_a''(x)\psi_a'(x)+\beta\sigma_a'(x)\psi_a''(x)\,,
\end{equation}
with $\sigma_a(x)\to \sign(x)$ when $a\to 0$, corresponding to the regularization of (2) in the main text. We will show that the resulting wave function is continuous at $0$. Integrating \eqref{eqdelta5} between $x_0$ and $x$, one has
\begin{equation}\label{newo}
\psi'_a(x)-\beta \sigma'_a(x)\psi_a'(x)=\psi'_a(x_0)-\beta \sigma'_a(x_0)\psi_a'(x_0)-k^2\int_{x_0}^x \psi_a(y)\D{y}\,.
\end{equation}
 Let us now specify
 \begin{equation}
\sigma'_a(x)=\frac{1}{2a}\pmb{1}_{|x|<a}\,.
\end{equation}
Integrating between $-a$ and $a$ one finds
\begin{equation}
(\psi_a(a)-\psi_a(-a))(1-\tfrac{\beta}{2a})=\mathcal{O}(a)\,,
\end{equation}
which imposes that
\begin{equation}\label{means}
\underset{a\to 0}{\lim}( \psi_a(a)-\psi_a(-a))=0\,.
\end{equation}
Now, integrating \eqref{newo} between $a$ and $a+\epsilon$ for $\epsilon>0$ we find
\begin{equation}\label{means2}
\psi_a(a+\epsilon)-\psi_a(a)=\mathcal{O}(\epsilon,a)\,.
\end{equation}
Defining
\begin{equation}
\psi(x)=\underset{a\to 0}{\lim}\,\psi_a(x)\,,
\end{equation}
we find that \eqref{means2} implies
\begin{equation}
\underset{a\to 0}{\lim}\,\psi_a(a)=\psi(0^+)\,.
\end{equation}
Similarly one has
\begin{equation}
\underset{a\to 0}{\lim}\,\psi_a(-a)=\psi(0^-)\,.
\end{equation}
Hence \eqref{means} means
\begin{equation}
\psi(0^+)=\psi(0^-)\ ,
\end{equation}
and the function is continuous at zero.

An important comment about this result is in order. The reader could have noticed that by solving \eqref{eqdelta5} perturbatively in $\beta$, one finds at order $\beta$ that the solution has a discontinuity compatible with the problem (4) in the main text. In fact, a discontinuity can be present perturbatively and absent non-perturbatively. The simplest differential equation with this behaviour is
\begin{equation}
\partial_x^2\psi_a-\beta \partial_x(\sigma_a' \partial_x\psi_a)=0\,.
\end{equation}
The solution at first order in $\beta$ is \begin{equation}\label{pert}
\psi_a(x)=A(x+\beta\sigma_a(x))+B+\mathcal{O}(\beta^2)\,,
\end{equation}
with $A,B$ integration constants, which has indeed a discontinuity at zero in the limit $a\to 0$. However, the exact solution is
\begin{equation}
\psi_a(x)=A'\int_{-1}^x \frac{\D{y}}{1-\beta\sigma_a'(y)}+B'\,,
\end{equation}
which is equal to $(x+1)A'+B'$ in the limit $a\to 0$, without any discontinuity.

\section{Feynman Rules}
\label{sec:FR}

In this section we report the Feynman rules for the diagrammatic calculations in the main text. The propagator line reads as  
\begin{fmffile}{rules}
\be
\parbox{30mm}{\begin{fmfgraph*}(80,50)
   \fmfleft{i}
   \fmfright{o}
   \fmf{fermion,label=$\omega_n,,k$}{i,o}
\end{fmfgraph*}}\qquad
=\frac{1}{i \omega_n -k^2+\mu}~.
\ee
Instead, the vertex is given by 
\be
\parbox{45mm}{\begin{fmfgraph*}(120,75)
\fmfleft{i1,i2}
\fmfright{o1,o2}
\fmf{fermion,label=${\omega_{n_4}},,{k_4}$, l.side=left}{i1,v1}
\fmf{fermion,label=${\omega_{n_1}},,{k_1}$, l.side=left}{v1,o1}
\fmf{fermion,label=${\omega_{n_3}},,{k_3}$, l.side=right}{i2,v1}
\fmf{fermion,label=${\omega_{n_2}},,{k_2}$, l.side=right}{v1,o2}
\end{fmfgraph*}}
\qquad= 4  \delta_{\omega_{n_1}+\omega_{n_2}-\omega_{n_3}-\omega_{n_4},0}W_{a,\beta}(k_1,k_2,k_3,k_4) .
\ee
\end{fmffile}
In addition, on each internal line there is a sum over $k$ and $\omega$. Finally, the contribution of every diagram is multiplied by $(-1)^F S^{-1}$, where $F$ is the number of closed loops and $S$ is the symmetry factor, i.e. the number of ways in which internal lines can be exchanged whilst leaving the diagram invariant.

\section{Self Energy}
\label{sec:SR}

Evaluating the diagrams in Eq. (26) of the main text using the rules described in the previous section we find 
\begin{align}
\Sigma(\omega_n, p) =& \frac{1}{2}\Sigma_1(p)-\frac{1}{T} \sum_{q}\Sigma_{1}(q) W_{a,\beta}(p,q,q,p)n(q) (1-n(q))\notag\\
&- 2 \!\!\!\!\!\!\!\! \sum_{\substack{p_2, p_3 \\ p_4=p+p_2-p_3}}\!\!\!\!\!\!\!\!W_{a,\beta}(p,p_2,p_3,p_4)W_{a,\beta}(p_4, p_3,p_2,p) n(p_2) (1-n(p_3))(1-n(p_4)) f_T(i \omega_n +p_2^2-p_3^2-p_4^2+\mu),
\label{eq:selfenergy}
\end{align}
where $p_4=p+p_2-p_3$, $\omega_n =  {2\pi}(n+1/2) T$, and we introduced 
\begin{align}
\Sigma_{1}(p) &:= {4} \sum_{q} W_{a,\beta}(p,q,q,p)n(q)\,, &
f_T(x) &:=\frac{e^{x/T}-1}{x}.
\end{align}
Considering now the Fourier transform of the potential (cf. Eq.(6) of the main text)
\be
\hat V_{a,\beta}(k) = \frac{\beta}{a^2} \int {\rm d}x\,\, e^{i k x} \frac{\sigma''(x/a)}{x+\beta \sigma(x/a)}  
\label{eq:Etiennespotential}
\ee
and expanding in $\beta$ we find 
\be
\hat V_{a,\beta}(k)-\hat V_{a,\beta}(0) = \frac{F(ka)}{a^2}\beta+  \frac{G(ka)}{a^3} \beta^2 +O\left(\beta^2\right)\!,
\label{eq:Vexp}
\ee
where we defined 
\begin{align}
F(k) &:= \int {\rm d}x\,\, (e^{i k x}-1) \frac{\sigma''(x)}{x} =  \int^{k}_0 \!\!\!{\rm d}x\, x \hat \sigma'(x)\label{eq:Fdef}\\ 
G(k) &:= - \int {\rm d}x\,\, (e^{i k x}-1) \frac{\sigma''(x)\sigma(x)}{x^2}.
\end{align}
Finally, for small $k$ we have 
\be
F(k) = k^2+O(k^4),\qquad G(k) = - \frac{k^2}{2}  \int {\rm d}x\,\, \sigma'(x)^2 +O(k^4). 
\ee
Plugging \eqref{eq:Vexp} into (24) of the main text and taking the thermodynamic limit we have 
\be
\Sigma_{1}(p) = \Bigl[\beta-\frac{\beta^2}{2 a} \int {{\rm d} p}\,  \sigma'(p)^2 \Bigr] \tilde \Sigma_1(p)+O(a),
\label{eq:divergentfirstorder}
\ee
with 
\be
\tilde \Sigma_{1}(q) : =- 2 \int \frac{{\rm d} p}{2\pi}\, (p-q)^2  n(p) = - 2 A_2 -2 A_0 q^2,
\ee
where $A_m$ is defined in Eqs. (28) of the main text. 

The only other divergent term in \eqref{eq:selfenergy} is found considering the following contribution to the second line 
\be
\frac{2\beta^2 }{a^4} \int \frac{{\rm d} p_2}{2\pi}\frac{{\rm d} p_3}{2\pi}  \frac{(F(a(p-p_3))-F(a(p_2-p_3)))^2}{i\omega_n +p_2^2-p_3^2-(p+p_2-p_3)^2+\mu} n(p_2) 
\ee 
where we took the thermodynamic limit. To treat this term we deform the integration over $p_3$ to a path in the complex plane that is parallel to the real axis but shifted by an amount $i \eta/a$ for some $\eta>0$. To do that we assume that $ \tilde \sigma'(p)$ is holomorphic in a strip of width $\eta$ around the real axis, implying that also $F(p)$ is (cf.~\eqref{eq:Fdef}). In deforming the contour we pick a pole at
\be
p_+ = \frac{1}{2}(p+p_2)+\frac{1}{2}\sqrt{2(i \omega_n-p^2+\mu)+(p-p_2)^2}
\ee
where we took the branch cut of the square root along the positive real axis such that ${\rm Im}(p_+)>0$. In summary we have  
\begin{align}
&\frac{2\beta^2 }{a^4}\int \frac{{\rm d} p_2}{2\pi}\frac{{\rm d} p_3}{2\pi}  \frac{(F(a(p-p_3))-F(a(p_2-p_3)))^2 n(p_2)}{i\omega_n +p_2^2-p_3^2-(p+p_2-p_3)^2+\mu}   = \notag\\
&\frac{2\beta^2 }{a^4} \int \frac{{\rm d} p_2}{2\pi}\frac{{\rm d} p_3}{2\pi}  \frac{(F(a(p-p_3)-i \eta)-F(a(p_2-p_3)-i \eta))^2  n(p_2)}{i\omega_n +p_2^2-(p_3+i \eta/a)^2-(p+p_2-p_3-i \eta/a)^2+\mu}\notag\\
&-\frac{i \beta^2 }{a^4} \int \frac{{\rm d} p_2}{2\pi}  \frac{(F(a(p-p_+))-F(a(p_2-p_+)))^2 n(p_2)}{(p_+-p_-)} 
\end{align}
where we defined 
\be
p_- = \frac{1}{2}(p+p_2)-\frac{1}{2}\sqrt{2(i \omega_n-p^2+\mu)+(p-p_2)^2}\,.
\ee
Let us consider the two terms on the r.h.s. separately. Looking at the first term and expanding in $a$ we have  
\begin{align}
&\frac{2\beta^2 }{a^4}\int \frac{{\rm d} p_2}{2\pi}\frac{{\rm d} p_3}{2\pi}  \frac{(F(a(p-p_3)-i \eta)-F(a(p_2-p_3)-i \eta))^2  n(p_2)}{i\omega_n +p_2^2-(p_3+i \eta/a)^2-(p+p_2-p_3-i \eta/a)^2+\mu} \notag\\
&= -\frac{\beta^2 }{a} \int \frac{{\rm d} p_2}{2\pi}\frac{{\rm d} k}{2\pi}  \frac{(F'(k-i \eta))^2 (p_2-p)^2  n(p_2)}{(k-i \eta)^2} + O(a)\notag\\
&= -\frac{\beta^2 }{a} \int \frac{{\rm d} k}{2\pi} (\tilde\sigma'(k-i \eta))^2 \int \frac{{\rm d} p_2}{2\pi} (p_2-p)^2  n(p_2)+O(a) \notag\\
&= \frac{\beta^2 }{2 a} \left[\int {{\rm d} k}\, \sigma'(k)^2\right] \tilde \Sigma_1(p)  +O(a)\,.
\end{align}
In particular: the $O(a^0)$ term vanishes as one can verify using the identity
\be
\int {\rm d} x\, \frac{f'(x)f''(x)}{x^2} = \int {\rm d} x\, \frac{f'(x)^2}{x^3}\,.
\ee
Instead, the second term gives
\begin{align}
&-\frac{i \beta^2}{a^4} \int \frac{{\rm d} p_2}{2\pi}  \frac{(F(a(p-p_+))-F(a(p_2-p_+)))^2 n(p_2)}{(p_+-p_-)}\notag\\
&=- i \beta^2 \int \frac{{\rm d} p_2}{2\pi}  \frac{((p-p_+)^2-(p_2-p_+)^2)^2 n(p_2)}{(p_+-p_-)} \notag\\
&=- i \beta^2 \int \frac{{\rm d} p_2}{2\pi}  \sqrt{2(i \omega_n-p^2+\mu)+(p-p_2)^2} (p-p_2)^2 n(p_2), 
\end{align}
where in the second term we took the limit $a\to0$. Putting together the finite terms we find  
\begin{align}
\Sigma(\omega_n, p) =& -{2\beta}(A_2 + A_0 p^2)-\frac{2\beta^2}{T} \int\frac{{\rm d} q}{2\pi} (A_2 + A_0 q^2) (p-q)^2 n(q) (1-n(q))\notag\\
&+2 \beta^2 \int \frac{{\rm d} p_2}{2\pi}\frac{{\rm d} p_3}{2\pi} {((p-p_3)^2-(p_2-p_3)^2)^2}   \frac{(n(p_3)n(p_4)-n(p_2)n(p_3)-n(p_2)n(p_4))}{i\omega_n +p_2^2-p_3^2-p_4^2+\mu}\notag\\
&-i \beta^2 \int \frac{{\rm d} p_2}{2\pi}  \sqrt{2(i \omega_n-p^2+\mu)+(p-p_2)^2} (p-p_2)^2 n(p_2)\,.
\label{eq:SigmaR}
\end{align}

\end{document}